\documentclass[a4paper,11pt]{article}
\usepackage{pos}

\title{Test of lepton family universality and search for lepton and baryon number violation at Belle}

\author*[a,b]{D. Sahoo}

\affiliation[a]{Tata Institute of Fundamental Research, Mumbai, India}

\affiliation[b]{Utkal University, Bhubaneswar, India \\

On behalf of the Belle Collaboration}

\emailAdd{sahoodev1994@gmail.com}

\abstract{The electroweak penguin B decays mediated by $b \to s \ell^+ \ell^- $ transitions are flavour-changing neutral current processes, and are thus sensitive to new physics owing to potential contributions of heavy particles in the quantum loop.
Recently, LHCb has obtained interesting results, where the possible hints of lepton family universality violation (LFUV) could be seen.
We report a new measurement of the LFUV observable $R_{K}$, the ratio of branching fractions of $B\to K\mu^{+}\mu^{-}$ to $B\to Ke^{+}e^{-}$, based on the full data sample recorded by Belle at the $\Upsilon(4S)$ resonance from
$e^+e^-$ collisions produced by the KEKB collider.
We also report results on lepton flavor violating $B$ decays, $B^{+}\to K^{+}\mu^{\pm}e^{\mp}$ and $B^{0}\to K_{S}^{0}\mu^{\pm}e^{\mp}$.
The $B$-factory at KEK is also a $\tau$ factory, creating a copious amount of $\tau\tau$ pairs.
We have used these data to look for lepton and baryon number violating $ \tau$ decays  $\tau^{-}\to\overline{p}e^{+}e^{-}$, $pe^{-}e^{-}$, $\overline{p}e^{+}\mu^{-}$, $\overline{p}e^{-}\mu^{+}$, $\overline{p}\mu^{+}\mu^{-}$, and $p\mu^{-}\mu^{-}$.
Results of the search are also reported.}

\FullConference{%
  40th International Conference on High Energy physics - ICHEP2020\\
  July 28 - August 6, 2020\\
  Prague, Czech Republic (virtual meeting)
}

\begin{document}
\maketitle

\section{KEKB and Belle}

The Belle detector~\cite{belle} was placed at the interaction point of the KEKB asymmetric-energy $e^{+}e^{-}$ collider~\cite{kekb}.
It was a large-solid-angle magnetic spectrometer comprising six subdetectors: silicon vertex detector, central drift chamber, aerogel Cherenkov counter, time-of-flight counter, CsI(Tl) crystal electromagnetic calorimeter, and $K^{0}_{L}$ and muon detector.
The data used in the studies reported here were recorded by Belle at and near the $\Upsilon(4S)$ and $\Upsilon(5S)$ resonances.  

\section{Test of lepton family universality}

In the standard model (SM), the electroweak couplings of gauge bosons to leptons are independent of their flavor and the property is known as lepton family universality.
The decays $B\to K\ell^{+}\ell^{-} (\ell =e,\mu)$, mediated by the $b\to s\ell^{+}\ell^{-}$ quark-level transition, constitute flavor-changing neutral current processes.
Such processes are forbidden at tree level in the SM but can proceed via suppressed loop-level diagrams, and are therefore sensitive to particles predicted in a number of new physics models~\cite{bmodel1, bmodel2}. 
A robust observable to test the SM prediction is the lepton-family-universality (LFU) ratio
\begin{equation*}
    {R_{K}=\frac{\int{\frac{d\Gamma}{dq^{2}}}[B\to K\mu^{+}\mu^{-}]\,dq^{2}}{\int{\frac{d\Gamma}{dq^2}}[B\to Ke^{+}e^{-}] \,dq^{2}}},
\end{equation*}
where the decay rate $\Gamma$ is integrated over the available range of the dilepton invariant-mass squared, $q^{2}\equiv  M^2(\ell^+\ell^-)$.
LHCb \cite{rklhcb} has measured $R_{K}$ with a difference of about 2.5 standard deviations $(\sigma)$ from the SM prediction in the $q^{2}\in (1.1, 6.0)$ ${\rm GeV^2}$ bin.
Previous measurement of the same quantity was performed by Belle in the whole $q^{2}$ range using a data sample of $657\times 10^{6} B\overline{B}$ events.

\begin{figure}
\centering
\includegraphics[width=.55\linewidth]{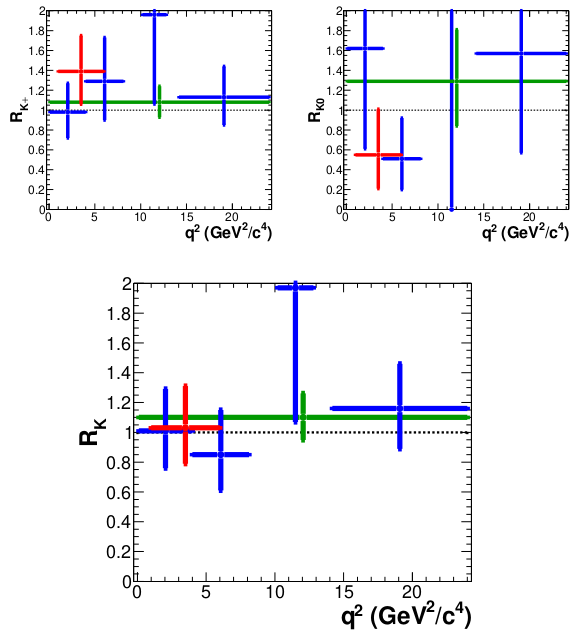}
\caption{$R_{K}$ in bins of $q^2$ for $B\to K^{+}\ell^{+}\ell^{-}$ (top left), $B^{0}\to K_{S}^{0}\ell^{+}\ell^{-}$ (top right), and both modes combined (bottom). The red marker represents the bin of $1<q^{2}<6$ $\mathrm{GeV^2}$, and blue markers are for $0.1<q^{2}<4$, $4<q^{2}<8.12$, $10.2<q^{2}<12.8$ and $q^{2}>14.18$ $\mathrm{GeV^2}$ bins. The green marker is for the whole $q^2$ region excluding the charmonium resonances. }
\label{fig:rk}
\end{figure}

We present here the results~\cite{rkb} obtained from a multidimensional fit performed on the full $\Upsilon(4S)$ data sample of Belle, which replace our previous result~\cite{rkBelle}.
Following four channels are studied:
$B^{+}\to K^{+}e^{+}e^{-}$, $B^{+}\to K^{+}\mu^{+}\mu^{-}$, $B^{0}\to K_{S}^{0}e^{+}e^{-}$, and $B^{0}\to K_{S}^{0}\mu^{+}\mu^{-}$ based on 711 $\rm fb^{-1}$ $\Upsilon(4S)$ data corresponding to $772\times 10^{6}$ $B\overline{B}$ events.  The signal yield is extracted with an extended unbinned maximum-likelihood fit to the distributions of $M_{\rm bc}\equiv\sqrt{E_{\rm beam}^{2}-\lvert\vec{p}_{B}^{\,2}\rvert}$, $\Delta E\equiv E_{B}-E_{\rm beam}$, and neural network output (background suppression variable).
Here $E_{\rm beam}$ is the beam energy, and $E_{B}$ and $\vec{p}_{B}$ are the energy and momentum of the reconstructed $B$ meson candidates, all calculated in the center-of-mass (CM) frame.
From the fit we obtain $137\pm 14$ and $138\pm 15$ events in $B^{+}\to K^{+}\mu^{+}\mu^{-}$ and $B^{+}\to K^{+}e^{+}e^{-}$ decays, respectively.
Similarly, the yields for the neutral channels $B^{0}\to K_{S}^{0}\mu^{+}\mu^{-}$ and $B^{0}\to K_{S}^{0}e^{+}e^{-}$ are $27.3_{\,-\,5.8}^{\,+\,6.6} $ and $21.8_{\,-\,6.1}^{\,+\,7.0} $ events.
The fit has also been performed in different $q^{2}$ bins and the $R_K$ results are shown in Figure~\ref{fig:rk}.
The $R_{K}$ values for different $q^{2}$ bins are consistent with the SM predictions, and the value for the whole $q^{2}$ range is $1.10^{\,+\,0.16}_{\,-\,0.15} \pm 0.02$.
Our $R_{K^{+}}$ result for $q^{2}\in (1.0,6.0)\,\rm{GeV^2}$ is higher by 1.6$\sigma$ compared to the LHCb result. 

\section{Lepton flavor violation in $B$ decays} 

In many theoretical models, lepton flavor violation (LFV) accompanies LFU violation~\cite{accompany}.
Extrapolating from the level of neutrino mixing, LFV is only possible at rates far below the current experimental sensitivity.
In case of a signal, this will unequivocally constitute signature of physics beyond the SM.
The LFV in $B$ decays can be studied via $B\to K\mu^{\pm}e^{\mp}$.
The most stringent upper limits on $B^{+}\to K^{+}\mu^{+}e^{-}$ and $B^{+}\to K^{+}\mu^{-}e^{+}$ set by LHCb~\cite{blfv} are $6.4\times 10^{-9}$ and $7.0\times 10^{-9}$ at 90$\%$ confidence level (CL).
Also, $B^{0}\to K^{0}\mu^{\pm}e^{\mp}$ decays were searched for by BaBar~\cite{babrlfv}, setting a 90$\%$ CL upper limit on their branching fractions at $2.7\times 10^{-7}$.   

\begin{figure}
\centering
\includegraphics[width=.7\linewidth]{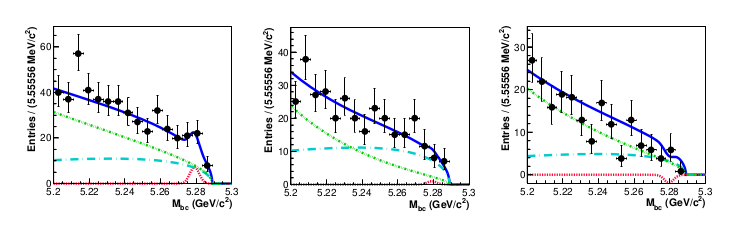}
\caption{Signal-enhanced $M_{\rm bc}$ projection of three-dimensional unbinned maximum likelihood fits to the data for decays $B^{+}\to K^{+}\mu^{+}e^{-}$ (left), $B^{+}\to K^{+}\mu^{-}e^{+}$ (middle), and  $B^{0}\to K_{S}^{0}\mu^{\pm}e^{\mp}$, respectively. Points with error bars are the data, blue solid curves are the fitted results for the signal-plus-background hypothesis, red dashed curves denote
the signal component, while cyan long-dashed, green dash-dotted, and black dashed curves represent
continuum, $B\overline{B}$ background, and charmless $B$ decays, respectively.}
\label{fig:lfvb}
\end{figure}

The signal yields for LFV decays in Belle~\cite{rkb} are obtained by performing extended maximum-likelihood fits, similar to those for $B\to K\ell^{+}\ell^{-}$ channels.
The signal-enhanced projection plots obtained from the fit for LFV decays are shown in Figure ~\ref{fig:lfvb}.
The fitted yields are $11.6^{\,+\,6.1}_{\,-\,5.5}$, $1.7^{\,+\,3.6}_{\,-\,2.2}$, and $-3.3^{\,+\,4.0}_{\,-\,2.8}$ for $B^{+}\to K^{+}\mu^{+}e^{-}$, $B^{+}\to K^{+}\mu^{-}e^{+}$, and $B^{0}\to K_{S}^{0}\mu^{\pm}e^{\mp}$, respectively.
The 90$\%$ CL upper limits on the branching fractions are ${\cal B}(B^{+}\to K^{+}\mu^{+}e^{-})<8.5\times 10^{-8}$, ${\cal B}(B^{+}\to K^{+}\mu^{-}e^{+})<3.0\times 10^{-8}$, and ${\cal B}(B^0 \to K_{s}^{0}\mu^{\pm}e^{\mp}) < 1.9 \times 10^{-8}$.
As there is a $3.2\sigma$ evidence for signal in $B^{+}\to K^{+}\mu^{+}e^{-}$, we also quote the branching fraction of $(4.98_{\,-\,2.36}^{\,+\,2.62}\pm 0.13)\times 10^{-8}$.
The earlier limit on the neutral decay channel is improved by an order of magnitude. 

\section{Lepton-flavor-, lepton-number- and baryon-number-violating $\tau$ decays}

As lepton flavor, lepton number (\textit{L}) and baryon number (\textit{B}) are accidental symmetries of the SM, there is no reason to expect them to be conserved in all possible particle interactions.
In fact, lepton flavor violation has already been observed in neutrino oscillations~\cite{oscl}.
While \textit{B} is presumed to have been violated in the early universe, its exact mechanism still remains unknown.
It is one of the three criteria formulated by Sakharov~\cite{sakharov} to explain the matter-antimatter asymmetry observed in the universe.
Any observation of processes involving \textit{B} violation would be a clear signal of new physics.
Such processes are studied in different scenarios of physics beyond the SM, such as supersymmetry~\cite{susy}, grand unification~\cite{grnd}, and models with black holes~\cite{black}.

Based on $1\,{\rm fb}^{-1}$ of $pp$ collision data, LHCb~\cite{lhcb} has studied the last two channels, setting $90\%$ CL upper limits on their branching fractions: ${\cal B}(\tau^{-}\to\overline{p}\mu^{+}\mu^{-})<3.3\times10^{-7}$ and ${\cal B}(\tau^{-}\to p\mu^{-}\mu^{-})<4.4\times10^{-7}$.
Using experimental bounds on proton decay, authors in Refs.~\cite{theory1,theory2,theory3} predict a branching fraction in the range of $10^{-30}$--$10^{-48}$ for these kinds of decays. We report herein the results~\cite{taubn} on a search for six \textit{L}- and \textit{B}-violating decays: $\tau^{-}\to\overline{p}e^{+}e^{-}$, $pe^{-}e^{-}$, $\overline{p}e^{+}\mu^{-}$, $\overline{p}e^{-}\mu^{+}$, $\overline{p}\mu^{+}\mu^{-}$, and $p\mu^{-}\mu^{-}$~\cite{charge} using 921 fb$^{-1}$ of data, equivalent to $(841\pm12)\times 10^6$ $\tau^{+}\tau^{-}$ events, recorded with the Belle detector at the KEKB asymmetric-energy $e^{+}e^{-}$ collider. 

The $\tau$ lepton is reconstructed by combining a proton or an antiproton with two charged lepton candidates.
To identify the signal, we use two kinematic variables: the reconstructed mass $M_{\rm rec}\equiv\sqrt{E_{p\ell\ell^{\prime}}^{2}-\vec{p}_{p\ell\ell^{\prime}}^{\,2}}$ and the energy difference $\Delta E\equiv E^{\rm CM}_{p\ell\ell^{\prime}}-E^{\rm CM}_{\rm beam}$, where $E_{p\ell\ell^{\prime}}$ and $\vec{p}_{p\ell\ell^{\prime}}$ are the sum of energies and momenta, respectively, of the $p$, $\ell$ and $\ell^{\prime}$ candidates.
The beam energy $E^{\rm CM}_{\rm beam}$ and $E^{\rm CM}_{p\ell\ell^{\prime}}$ are calculated in the CM frame. 

To optimize the event selection and obtain signal detection efficiency, we use Monte Carlo (MC) simulation samples. Various MC generators are used to generate background and signal MC samples.
Background samples include $e^{+}e^{-}\to\tau^{+}\tau^{-}(\gamma)$,  $e^{+}e^{-}\to q\overline{q}$ ($udsc$ continuum and $B\overline{B}$), Bhabha scattering, dimuon and two-photon mediated events. 

\begin{figure}[!htb]
\begin{center}
 \includegraphics[width=0.65\textwidth]{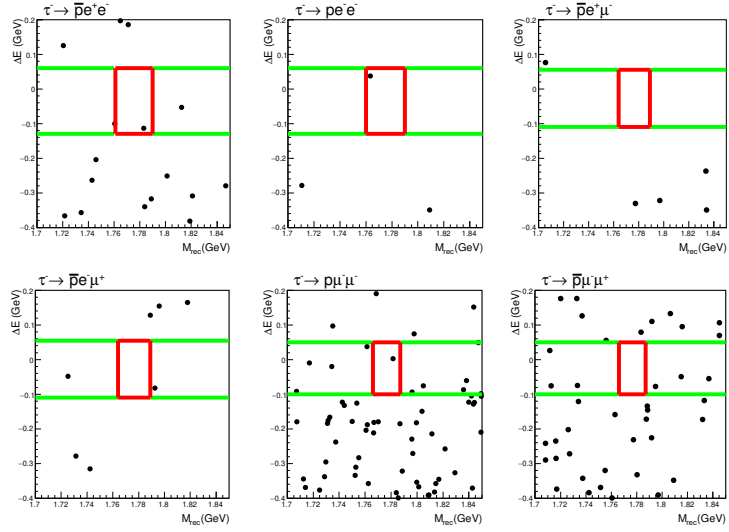}
\end{center}
\caption{$\Delta E$--$M_{\rm rec}$ distribution where the red box denotes the signal region and the green $\Delta E$ strip is used to calculate the expected background. Black dots represent the data.}
\label{fig:fig1}
\end{figure}

At the preliminary level, we try to retain as much generic $e^{+}e^{-}\to\tau^{+}\tau^{-}$ events as possible in the sample while reducing obvious backgrounds by applying suitable selection requirements on different kinematic variables.
At the next stage of selection, we apply dedicated selection criteria to pick up candidate events that are more signal-like.
We perform a sideband study to identify the sources of background that are dominated by events with a misidentified proton or antiproton, as well as to verify the overall data-MC agreement.
Since we follow the blind analysis method, before looking at data in the signal region, we estimate the background contribution in that region.
For this we choose a $\Delta E$ strip by hiding the signal region to predict the background expected in that region as shown in Figure~\ref{fig:fig1}. 

We don't find any excess of events in the signal region as shown in Figure~\ref{fig:fig1} and Table~\ref{tab:tab1}.
As the number of events observed is consistent with the background prediction, we calculate an upper limit on the signal yield using a frequentist method based on a double-sided profile likelihood test-statistic~\cite{ul1,ul2} with $\rm {CL_{s+b}}$ as the p-value.
We have set $90\%$ CL upper limits on the branching fractions of these lepton flavor, lepton and baryon number violating tau decays in the range of $(1.9$--$3.9)\times 10^{-8}$.
In Table~\ref{tab:tab1} we list results for all channels. In the case of $\tau^{-}\to p\mu^{-}\mu^{-}$ and $\overline{p}\mu^{-}\mu^{+}$, our limits are improved by an order of magnitude compared to LHCb~\cite{lhcb}.
For the remaining four channels, we set limits for the first time.

\begin{table}[h]
\caption{Signal detection efficiency ($\epsilon$), number of expected background events ($N_{\rm bkg}$), number of observed data events ($N_{\rm obs}$), 90$\%$ CL upper limits on the signal yield and branching fraction for various decay channels.}
\label{tab:tab1}
\begin{center}
\begin{tabular}{cccccc}
\hline\hline
Channel&$\epsilon\,(\%)$&$N_{\rm bkg} $&$N_{\rm obs}$&$N_{\rm sig}^{\rm UL}$&${\cal B}\,(\times 10^{-8})$\\
\hline
$\tau^{-}\to\overline{p}e^{+}e^{-}$     & $7.8$ & $0.50\pm 0.71$ & $1$ & $3.3$ & $<2.5$\\
$\tau^{-}\to pe^{-}e^{-}$               & $8.0$ & $0.21\pm 0.46$ & $1$ & $3.8$ & $<2.8$\\
$\tau^{-}\to\overline{p}e^{+}\mu^{-}$   & $6.5$ & $0.20\pm 0.44$ & $0$ & $2.3$ & $<2.1$\\ 
$\tau^{-}\to\overline{p}e^{-}\mu^{+}$   & $6.9$ & $0.40\pm 0.63$ & $0$ & $2.2$ & $<1.9$\\
$\tau^{-}\to p\mu^{-}\mu^{-}$           & $4.6$ & $1.30\pm 1.14$ & $1$ & $3.0$ & $<3.9$\\
$\tau^{-}\to\overline{p}\mu^{-}\mu^{+}$ & $5.0$ & $1.14\pm 1.07$ & $0$ & $2.3$ & $<2.7$\\
\hline
\end{tabular}
\end{center}
\end{table}

\section{Summary}

We have presented recent Belle results on $R_K$ and LFV searches.
We also report preliminary Belle results on \textit{L}- and \textit{B}-violating $\tau$ decays.
More precision and improved results on these studies are expected with its upgrade, namely Belle II.


\begin{thebibliography}{99}
\bibitem{charge} Inclusion of charge-conjugate processes is implied unless explicitly stated otherwise.
\end{thebibliography}
\end{document}